**AtomAI: A Deep Learning Framework for Analysis of Image and Spectroscopy Data in (Scanning) Transmission Electron Microscopy and Beyond**

Maxim Ziatdinov,[1,2,*] Ayana Ghosh,[1,2] Tommy Wong,[1,3] and Sergei V. Kalinin[1]

[1] The Center for Nanophase Materials Sciences and [2] Computational Sciences and Engineering Division, Oak Ridge National Laboratory, Oak Ridge, TN 3783, United States

[3] The Bredesen Center, University of Tennessee, Knoxville, United States

AtomAI is an open-source software package bridging instrument-specific Python libraries, deep learning, and simulation tools into a single ecosystem. AtomAI allows direct applications of the deep convolutional neural networks for atomic and mesoscopic image segmentation converting image and spectroscopy data into class-based local descriptors for downstream tasks such as statistical and graph analysis. For atomically-resolved imaging data, the output is types and positions of atomic species, with an option for subsequent refinement. AtomAI further allows the implementation of a broad range of image and spectrum analysis functions, including invariant variational autoencoders (VAEs). The latter consists of VAEs with rotational and (optionally) translational invariance for unsupervised and class-conditioned disentanglement of categorical and continuous data representations. In addition, AtomAI provides utilities for mapping structure-property relationships via *im2spec* and *spec2im* type of encoder-decoder models. Finally, AtomAI allows seamless connection to the first principles modeling with a Python interface, including molecular dynamics and density functional theory calculations on the inferred atomic position. While the majority of applications to date were based on atomically resolved electron microscopy, the flexibility of AtomAI allows straightforward extension towards the analysis of mesoscopic imaging data once the labels and feature identification workflows are established/available. The source code and example notebooks are available at https://github.com/pycroscopy/atomai.

---

[*] ziatdinovma@ornl.gov

# I. INTRODUCTION

Over the last decades, electron[1-9] and scanning probe microscopies[10-16] have emerged as keystone tools for the exploration of matter on the atomic and mesoscale levels. Applications of these techniques span the disciplines from fundamental condensed matter physics and materials science to biology and medicine, with the corresponding length scales ranging from atomic to micron and beyond. The recent exceptional progress in detectors and sensors, electron sources and computer-based data storage and analysis systems on the other, has opened the floodgates of high veracity structural and spectral data containing a wealth of information on materials structures and functionalities.

In (Scanning) Transmission Electron Microscopy, (S)TEM, the development of the aberration correction in the late 90's have led to a broad spectrum of experimental observations of atomic structures and functionalities.[3, 17, 18] The early successes of aberration corrected STEM included single-atom electron energy loss spectroscopy (EELS) imaging, allowing examination of the chemical state of a single impurity atom,[19] visualizations of the structures of a range of grain boundaries in broad materials classes,[20-22] that led to new fundamental insights into structural properties and high-temperature superconductivity, visualization of the light elements, and many others. Broad commercialization of aberration corrected STEMs has made these advances accessible in multiple research groups and has further stimulated the development of new STEM modalities enabled by further advances in electron sources and detectors. On the source side, these advances have begun to include beams with orbital angular momentum,[23-25] enabling probing of magnetic and orbital phenomena. The development of monochromators has improved the energy resolution in EELS to well below 10 meV, enabling direct probing of plasmons, phonons, and even the anti-Stokes excitations in complex materials.[26, 27] A combination of new detectors and monochromators has enabled momentum-resolved EELS measurements, effectively probing quasiparticle dispersion in $k$-space.[28, 29] Finally, the development of pixelated detectors has led to the broad introduction of 4D STEM, a method fundamentally based on detection of diffraction patterns from with in-plane localization below unit cell levels.[5, 30, 31] Advances in tomographic imaging and reconstruction have enabled 3D structural models of nanoparticles containing tens of thousands of atoms to be constructed with atomic resolution.[32]

The advances in STEM imaging and spectroscopy capabilities have naturally led the scientific community to explore new opportunities in probing physics and chemistry on the atomic scale. Until ~2010, the vast majority of (S)TEM applications were preponderantly qualitative in nature, where the observed atomic patterns, presence of specific defect types, or EELS responses were interpreted as indicators of certain *a priori* known physical behaviors or suggested emergence of new defect classes or structural elements. However, it was realized that the *quantitative* information on atomic positions can be used with EELS peak intensities to derive specific materials functionalities. As the first harbinger of these developments, Jia et al.[33, 34] and Chisholm et al.[35] demonstrated by TEM and STEM, respectively, that quantitative measurements of atomic column positions can be used to map the polarization order parameter field (a comprehensive account of polarisation measurements in ferroelectrics in the earlier days of atomic resolution imaging can be found in Ref [[36]]). This approach was rapidly extended to other physical functionalities strongly coupled to structure, including octahedra tilting in perovskites both in the image plane[37-39] and in the beam direction,[40, 41] chemical and physical strain fields,[42, 43] etc. These recent advances in quantitative STEM methods offer new opportunities for sub ~pm information mapping,[44] further allowing for direct connection to the generative physical models of solids.[45, 46]

Similar advances were explored in the scanning probe microscopy (SPM) community. While traditionally scanning tunnelling microscopy (STM) data is interpreted qualitatively, the structural mapping of atomic positions can yield information on strains[47] and crystal field splitting.[48] The spectroscopic measurements can be directly connected to the physical models[49,50] of solids or used for recognition imaging.[51]

Despite the markedly different imaging mechanisms, the nature of the detected signal, and its relationship to materials functionalities, STEM and SPM imaging exhibit commonalities both on the data format and physics extraction sides. In all cases, a scan gives rise to a scalar, multimodal, or hyperspectral data in the form of a structured N-dimensional array allowing for common analysis tools in the spectral domain.[52-55] Similarly, the objects of interest such as atomic positions in STEM, STM, and non-contact atomic force microscopy (AFM) are universal across these modalities. Most curiously, in many cases, the fundamentally different physics of imaging – convolution of the electron beam profile with delta-function like nuclei in STEM and convolution between the surface and probe density of states around the Fermi level in STM – gives rise to the superficially similar shapes of the objects as seen in images. Consequently, many atom finding or phase identification problems in STEM and STM can be performed with very similar analysis tools with minimal cross-technique modification.[56]

The purpose of the AtomAI package is to provide an environment that bridges the instrument specific libraries and general physical analysis (Fig. 1) by enabling the seamless deployment of machine learning (ML) algorithms including deep convolutional neural networks, invariant variational autoencoders, encoders-decoders, and decomposition/unmixing techniques for image and hyperspectral data analysis. Ultimately, it aims to combine the power and flexibility of the PyTorch deep learning framework[57] and simplicity and intuitive nature of packages such as scikit-learn,[58] with a focus on scientific data. To date, the majority of AtomAI applications have been for atomically resolved STEM.[59-62] However, as discussed above, many of its modules can equally be applied for atomically-resolved imaging in Scanning Tunneling Microscopy (STM)[63] and extended towards mesoscopic imaging in SPM[64] and STEM.[65]

## II. PYTHON BASED IMAGE ANALYSIS ECOSYSTEM

Recently, we have formulated a roadmap for the application of ML methods in imaging.[66] On the basic level, the questions we seek to answer are the following:

A. Can we get materials specific information (e.g., atomic coordinates from STEM, scattering potentials from 4D STEM, etc.) from microscopy data, for which the level of confidence, and how this knowledge is affected, can be improved from knowledge of the imaging system (e.g., classical beam parameters, resolution function, all the way to full imaging system modeling) and knowledge of a material's phase, structure, and composition?

B. Can we use the materials-specific information with uncertainties determined by incomplete knowledge of an imaging system or intrinsic physics limitations to infer physics and chemistry, either via correlative models or recovery of generative physics (force fields, exchange integrals, and other parameters)?

C. Can we use the inferred materials information, either correlative or causative, to reconstruct materials behavior (phase diagrams, etc.) in a broader parameter space (e.g., for temperatures and concentrations different for the specific sample studied) and determine how the reliability of such predictions depend on the position in the parameter space?

D. Can we harness the data streams from microscopes to engender real-time feedback, e.g., for autonomous experimentation and atomic manipulation?

Advances in (S)TEM and SPM imaging have stimulated the development of software ecosystems for data analytics allowing for the import (ingestion) of the data from instrumental formats, instrumental corrections, and simple analytics including multivariate statistics, addressing goal (A). Our goal for AtomAI is to interface A to C. Below, we briefly overview existing elements of the Python infrastructure for imaging and physics.

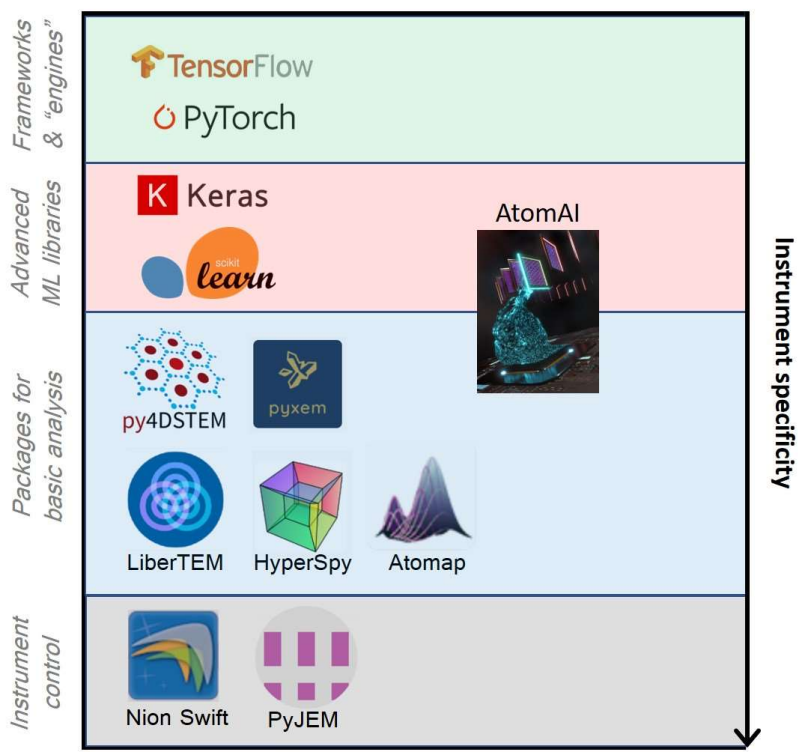

**Figure 1.** AtomAI is a flexible and user-friendly package for deep learning-based image analytics. It serves as a bridge between instrument specific libraries and general deep/machine learning frameworks. AtomAI logo is courtesy of O. Dyck.

One of the core libraries for reading, visualizing, and basic analysis of microscopy data is HyperSpy.[67] HyperSpy incorporates syntax for storing large multi-dimensional datasets, which facilitates the analysis of EELS, energy dispersive X-ray (EDX) spectra, and electron holography. It also incorporates a GUI to streamline the user experience. HyperSpy is currently transitioning from a single package to a multiplicity of specialized packages that rely on a single package ('hyperspy') for common infrastructure. However, although there are several packages specialized in image analysis, there is no deep learning capability. Another family of STEM analysis methods is the Pycroscopy eco-system,[68] which includes a STEMTools toolkit[69] and PyTEMLib library[70] for model-based quantification analysis. A comprehensive set of tools with a focus on electron diffraction data is available from the pyxem[71] library. In addition, there are 4D-STEM-specific analysis codes currently under development including LiberTEM,[72] and py4DSTEM.[73] LiberTEM is an open-source, GUI-based Python implementation that incorporates distributed computing to analyze large, multidimensional data. The core of LiberTEM is a framework for MapReduce-like operations on live data streams. All these packages are fully open source and can be freely modified to match the specific needs of a given research project. More information about these and some other (not open-sourced) packages can be found in the Supplementary Materials and Ref [74].

Complementary to the data analysis are methods for modeling STEM/EELS, including μSTEM,[75] QSTEM,[76] abTEM,[77] MULTEM,[78] STEMsalabim,[79] Prismatic,[80] Dr. Probe,[81] and others. These packages are used for simulating STEM images and EELS, as well as for convergent beam electron diffraction (CBED) calculations. Many of these packages can run efficiently on modern Graphics Processing Units (GPUs) allowing for a significant speed-up of the simulations and a viable source of data for training of ML models.

Finally, this description will be incomplete without mentioning the Python-based general physics simulation infrastructure. There exist ample resources/packages to perform simulations of physical systems at different length scales and for evaluating their properties. Some of these are open-source and various post-processing codes or scripts[82-93] are freely available for users. These include the atomic simulation environment (ASE), grid-based projector augmented wave (GPAW) package, and others.[94-103] With recent development in data analytics and machine learning capabilities useful for materials research, several databases[104-109] have also become popular allowing to perform simulations alongside data-driven studies. A non-exhaustive list of such contents along with their overall capabilities is listed in Supplementary Materials.

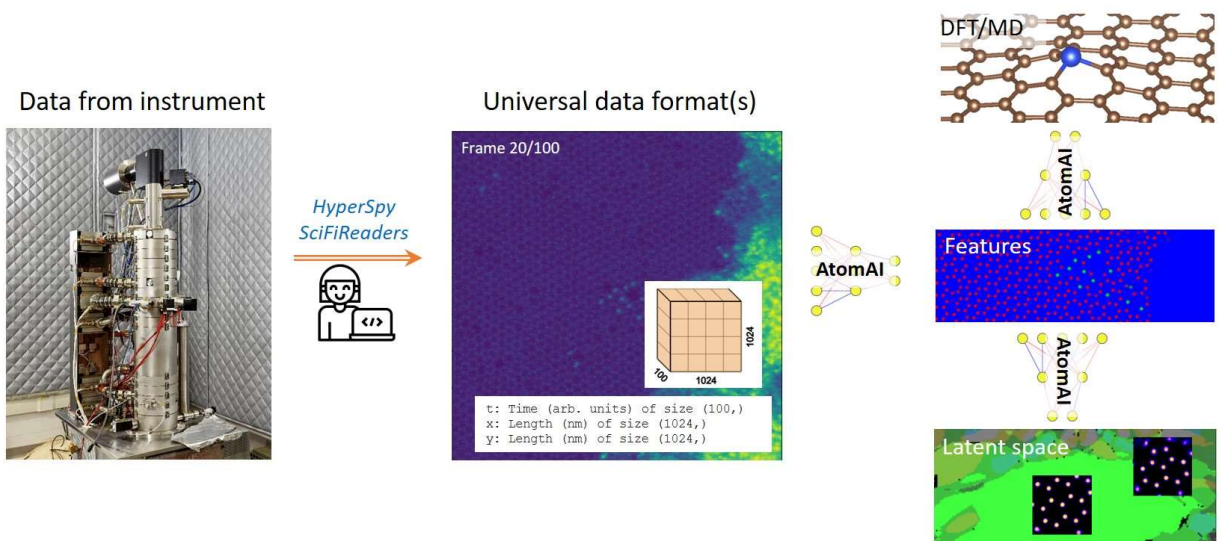

**Figure 2.** Schematic illustration of a typical AtomAI workflow. First, the microscopic images or spectra are transformed into an N-dimensional array object in NumPy (a fundamental package for scientific computing in Python). This can be achieved for example with a HyperSpy package. Since it is wise to keep the metadata (e.g., image acquisition parameters), one can convert experimental data into a Spectroscopic and Imaging (SID) data format using the SciFiReaders package.[110] AtomAI can work both with standard NumPy objects and SID objects. Once the data is in one of these formats, it can be passed through one of the AtomAI's pre-trained neural networks (or a new neural network can be trained if necessary) to extract features associated with atoms, defects, or other objects of interests (e.g., metallic nanoparticles or protein nanorods[111, 112]). Typically, features represent the type and position of atoms. The extracted features can then be used for more advanced analysis such as learning of disentangled factors of variation with deep latent space models[60] or they can serve as an input for first-principles calculations.[113] (The photo of a microscope is obtained from the ORNL photo gallery with permission for public release)

## III. PACKAGE CONTENT

AtomAI is an open-source Python library based on the PyTorch deep learning engine. The goal is to provide an environment bridging the instrument-specific libraries and general physical analysis by enabling seamless deployment of the deep and machine learning algorithms for image and hyperspectral data analytics. It aims to combine the power and flexibility of PyTorch deep learning framework and simplicity and intuitive nature of packages such as scikit-learn, with a focus on scientific image data. The Package content is briefly summarized in Table I.

**Table I**
1. Core modules (low-level API)
    a. Trainers
    b. Predictors
    c. Nets
2. AtomAI models (high-level API)
    a. Segmentor
    b. ImSpec
    c. Deep Ensembles
    d. Variational Autoencoders
3. Other utilities
    a. Multivariate Statistics
    b. Graph Analysis
    c. Atomic simulation environment
    d. Training data preparation

### III.A. Core modules and low-level API.

At the core of AtomAI's package are custom-built *trainers*, *predictors*, and neural *nets*.

The *trainers* wrap training routines for semantic segmentation models, im2spec models, variational autoencoder models, and deep ensembles. They incorporate some of the most recent advances in the deep learning field for model training including stochastic weight averaging, time-dependent weight perturbation, and on-the-fly data augmentation, which can be activated by simple Boolean statements (*e.g.*, `swa=True` turns stochastic weight averaging on). All the trainers also enforce a determinism in deep learning training such that different training runs lead to the same result (provided that models themselves do not contain sources of randomness). Note that while users don't have to interact with the *trainers* directly when using AtomAI models, they can use them for training custom models written in PyTorch with minimal lines of code.

The *predictors* wrap inference routines with trained models by automatically taking care of the proper dimensionality of the input data (e.g., adding a pseudo-dimension of 1 to the grayscale data) and possible size mismatch issues (by automatically performing data padding), as well as by allowing a batch-by-batch prediction on large datasets. Finally, both trainers and predictors automatically take care of all the necessary CPU-to-GPU (and vice versa) transfer of models and data.

The custom-built neural networks and associated building blocks are located in the *nets*. They include fully convolutional neural networks (FCNNs) for semantic segmentation including a U-Net[114] and a number of custom FCNNs such as:

i) DilNet, which uses only a single max-pooling operation to preserve the maximum amount of information while the utilization of cascades of the dilated convolutions instead of the regular convolutional blocks allows for the significant reduction of computational costs;

ii) ResHedNet, which allows for more accurate identification of edge features (fibers, domain walls, etc.) using a modified version of the holistically-nested edge detector[115] augmented with residual connections[116]; and

iii) SegResNet, which is a customized version of SegNet model[117] with residual convolutional blocks instead of the regular ones.

Another important subset of neural networks in the AtomAI's *nets* is encoder and decoder blocks used in models for predicting spectra from images (and vice versa) and for deep generative models. This includes both fully-connected ("MLP") and convolutional architectures of the encoders and decoders as well as architectures specific to the realization of rotationally and translationally invariant deep generative models. Several examples of the low-level API usage are given in the Supplementary Materials.

### III.B. AtomAI models and high-level API.

The *models* in AtomAI contain deep learning models for performing semantic segmentation of image data (Segmentor models), predicting spectra from images and vice versa (ImSpec), and a judicious collection of deep generative models based on variational autoencoders (VAE, *r*VAE, *j*VAE, *jr*VAE) with translational and/or rotational invariances for learning continuous and discrete latent representations of the data.

*Segmentor*. The semantic segmentation models categorize each pixel in the input image as belonging to a particular object (e.g., atom type, defect structure, etc.) or to a background. As such, it can be applied to problems such as atom finding in STM and STEM, identification of atomic steps in STM data, identification of domain walls in piezo-response force microscopy (PFM) data, and particle finding in atomic force microscopy (AFM) or optical microscope data. The semantic segmentation model can be initialized and trained with just a few lines of code:

```python
import atomai as aoi
# Initialize model
model = aoi.models.Segmentor(nb_classes=3)  # uses U-Net by default
# Train
model.fit(images, labels, images_test, labels_test, # training data
          training_cycles=300, compute_accuracy=True, swa=True # training params
          )
```

Here the `swa=True` argument turns on the stochastic weight averaging, which usually allows improving the model's accuracy and leads to better generalization.[118] The prediction with the trained Segmentor model requires just a single line of code:

```
nn_output, coordinates = model.predict(expdata)
```

Here, the coordinates are a dictionary with NumPy arrays of $N\times3$ dimension (keys correspond to image numbers in a stack) where $N$ is a number of detected objects (e.g., atoms), the first two columns correspond to the predicted $x$ and $y$ coordinates of the objects (e.g., atomic centers), and the third column corresponds to the predicted class (e.g., type of atom).

Note that there is also an option to refine the predicted position with a 2D Gaussian fit (by passing a `refine=True` keyword argument to model's `predict()` method) using the coordinates predicted by Segmentor as an initial guess.

*ImSpec*. In the *im2spec* model, the images (or image patches) representing a local structure are "compressed" via a convolutional neural network to a small number of latent variables, which are then "deconvoluted" to yield the spectra. In the *spec2im* model, the process is reversed. The *im2spec* models have been used for predicting hysteresis loops from topographic images in PFM experiments[119] and for predicting electron energy loss spectra from HAADF STEM images.[120] The initialization and training of the *im2spec* model are similar to that of the semantic segmentation model, with the main difference being that one has to specify the dimensions of input and output data:

```
in_dim = (16, 16)   # Input dimensions (image height and width)
out_dim = (64,)   # Output dimensions (spectra length)

# Initialize and train model
model = aoi.models.ImSpec(in_dim, out_dim, latent_dim=10)
model.fit(imgs_train, spectra_train, imgs_test, spectra_test,   # training data
          full_epoch=True, training_cycles=120, swa=True   # training parameters
          )
```

As with the semantic segmentation models, the prediction with the *im2spec* model takes a single line of code:

```
predicted_spectra = model.predict(imgdata)
```

*Deep Ensembles*. AtomAI can also be used to train ensembles of models. The mean ensemble prediction is usually more accurate and reliable than that of the single model.[121] In addition, it also yields information about the uncertainty in prediction for each pixel/point, which can further be used for anomaly detection[122] and implementation of automated experiment workflows.[123]

There are currently three strategies for training ensembles:

i) `train_ensemble_from_scratch` where each model in the ensemble starts with a different random initialization of weights and the data is shuffled differently for each model in the ensemble;

ii) `train_ensemble_from_baseline` where first a single model is trained for $N$ epochs and then used as a baseline to train an ensemble of models, each with the reset optimizer and different random shuffling of training data (guaranteeing a different training trajectory), for $n \ll N$ epochs;

iii) `train_swag` which performs sampling from a Gaussian subspace along a single training trajectory.

The ensemble training routines can be applied both to the built-in models and user-defined models. A code example of ensemble training can be found in the Supplementary Materials.

*Variational Autoencoders.* AtomAI has a built-in variational autoencoder (VAE)[124] and its multiple extensions for unsupervised determination of the most effective reduced representation of the system's local descriptors. Specifically, in addition to regular VAE, one can choose rotationally and (optionally) translationally invariant VAE (*r*VAE), as well as joint VAEs for disentangling continuous and discrete latent representations with (*jr*VAE) and without (*j*VAE) invariance to rotations and translations.

The VAEs can be applied to both raw data and output of a neural network, but typically work better with the latter (for example, passing raw data through a Segmentor prior to VAE analysis will typically lead to better convergence and improved results). Below is an example of initializing and training a *r*VAE which takes just a few lines of code:

```
# Get a stack of image patches from the experimental data or Segmentor output
window_size=32
imstack, com, frames = aoi.utils.extract_subimages(nn_output, coords, window_size)

# Initialize rVAE model
input_dim = (32, 32)
rvae = aoi.models.rVAE(input_dim)

# Train
rvae.fit(imstack_train, latent_dim=2, rotation_prior=np.pi/3,
         training_cycles=100, batch_size=100)
```

To visualize the learned latent manifold, one simply needs to run:

```
# Visualize the learned manifold
rvae.manifold2d();
```

### III.C. Other utilities.

AtomAI is not limited to deep learning applications and includes classes and methods for statistical and graph analysis, as well as utilities for training data preparation and conversion of deep learning predictions into the inputs for ab-initio simulations.

*Statistics.* AtomAI allows users to perform a multivariate statistical analysis on their datasets and/or on the predictions of the AtomAI's models. The statistics toolbox is available from the AtomAI's *stat* and includes principal and independent component analysis (PCA and ICA), non-negative matrix factorization (NMF), and Gaussian mixture model (GMM). In the example below, we use the *stat.imlocal* class to generate a stack of image patches centered around a specific class of objects predicted by the Segmentor (*e.g.*, a specific type of atom or defect) in a larger image and then perform one of the aforementioned types of statistical analysis:

```python
# Get local descriptors
imstack = aoi.stat.imlocal(nn_output, coords, window_size=32, coord_class=1)

# Compute distortion "eigenvectors" with associated loading maps
pca_results = imstack.imblock_pca(n_components=4, plot_results=True)
```

The *stat* also has functions for the refinement of atomic classes predicted by the Segmentor based on the statistical analysis of the intensities and local neighborhoods of the identified atomic features.

*Graph Analysis.* The AtomAI's *graphx* module can be used for the graph-based analysis of the atomic coordinates, typically from a Segmentor output. One of the applications is the identification of specific ring structures in coordinates data from carbon materials (e.g., graphene or nanotubes) using a depth-first search method. Because it constructs graphs using the information about actual atomic covalent radius, one has to specify a dictionary that will map classes from the Segmentor output (0, 1, …) into chemical elements (*e.g.*, 'C', 'Si', …). In addition, we do not assume that the metadata about scan size is always available (or correct) and hence a user needs to supply a coefficient for converting pixel coordinates to coordinates in angstroms.

*Atomic simulation environment.* AtomAI has two specific utility functions in the *aseutils*, namely `ase_obj_basic` and `ase_obj_adv` to convert the Segmentor-predicted coordinates into objects readable by the Python-based atomic simulations environment (ASE). These objects are also well-suited to be directly used in commonly known electronic structure codes such as VASP and can be visualized with packages such as VESTA.[97] The utility function `ase_obj_basic` reads in the dictionary containing the list of atomic positions and writes files assuming a cubic cell. Here, by default, the lattice parameters are assumed based on the maximum value of the coordinates. The `ase_obj_adv` gives a user an option to construct the cell. The user is asked to provide inputs for all lattice vectors along three dimensions.

*Training data preparation.* Finally, AtomAI's has utility functions for training data preparation, including generation of single-class and multi-class ground truth "masks" from atomic coordinates as well as data augmentation that can be performed before and/or during model training with both built-in functions (including blurring, Gaussian and Poisson noises, rotation, zooming, and resizing) and user-defined functions.

**IV. CASE STUDIES**

Here we illustrate several case studies using AtomAI. For more examples, please refer to the GitHub page of the project.[125]

**IV.A. Semantic segmentation of the atom-resolved images.**

Semantic segmentation of atomically resolved microscopic images involves classifying every pixel in the image as belonging to specific atom and/or defect classes. This is different compared to classifying natural images where we categorize one image as a whole. Because this is a supervised method, it requires training data where the atoms and/or defects are labelled. Once trained, a *Segmentor* model can be used to predict the position of the atoms and/or defects in previously unseen data. Below we show an example of how one can train a Segmentor model using labeled experimental images and apply the trained model to data from a different experiment alongside performing multivariate statistical analysis on the semantically segmented output.

To prepare the training data, we used a single labeled experimental STEM image from Sm-doped $BiFeO_3$ with the resolution of 3000 × 3000 pixels containing ~20,000 atomic unit cells.[122] The image is part of a publicly available dataset.[126] The corresponding mask (*aka* ground truth) was generated using the atomic coordinates from a Gaussian fit. Here, there are three different classes in the labelled data corresponding to atomic columns (hereafter referred to simply as "atoms") in A-lattice (center atom) and B-lattice (4 corner atoms). About 2000 patches of image-masks pairs with the size of 256 × 256 pixels were cropped and further "augmented" by applying different levels of noise, blurring, as well as changing scale (zooming-in) and 90°-rotations. The purpose of such augmentation is to account for variations in imaging conditions between different experiments. Figure 3 shows five different augmented images and corresponding ground truths.

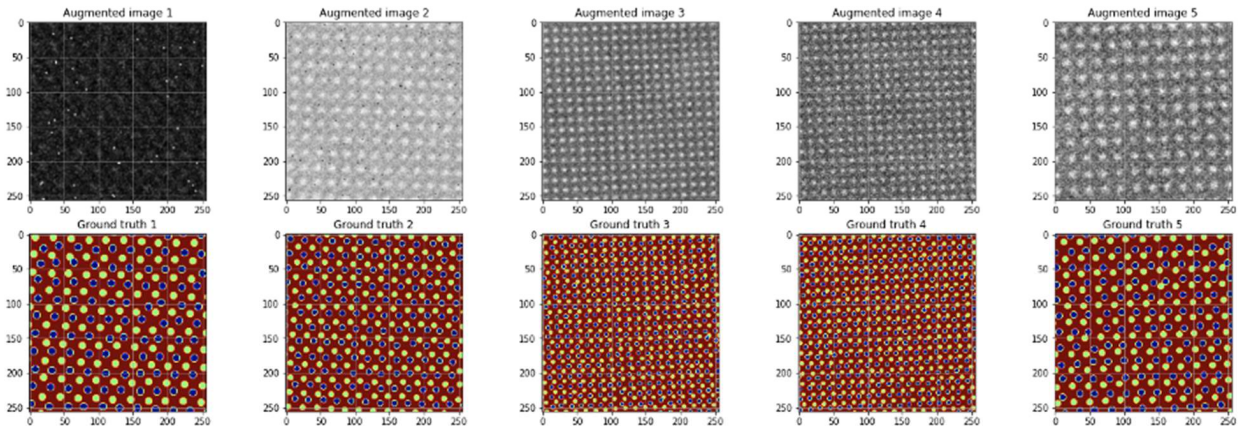

**Figure 3**. The top row represents the input to the Segmentor model and the bottom row represents the ground truth to which the output generated by the model will be compared during the model training.

The default Segmentor model is based on the U-Net neural network but one can also choose a different model or define a custom fully convolutional neural network. The Segmentor's raw output represents a set of well-defined (semantically segmented) blobs corresponding to different atomic types on a uniform background. For the trained model, the centers of the mass of the predicted blobs correspond to the atomic centers.

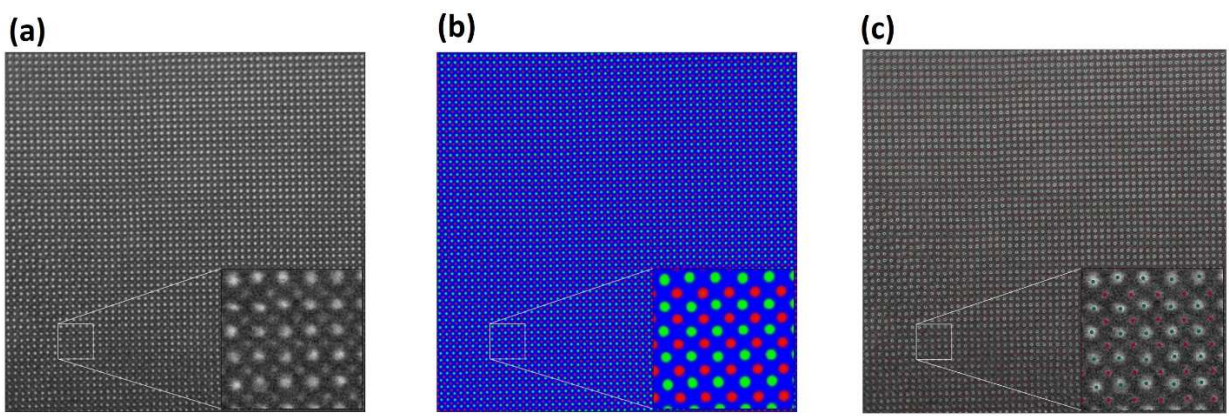

**Figure 4.** (a) STEM image from a similar material obtained in a different experiment (i.e., the Segmentor has not seen this image before). (b, c) Prediction of the trained Segmentor model: (b) semantically segmented raw output and (c) refined atomic coordinate (see the supplemental materials for a comparison between refined and non-refined prediction). Note that the model is robust with respect to variations in sample thickness.

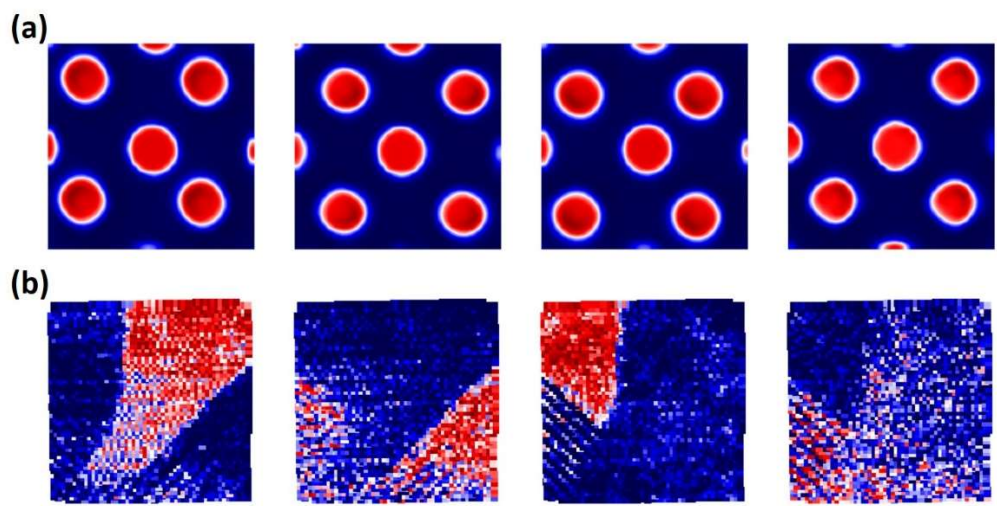

**Figure 5.** The results from the patch-based NMF analysis with the displacement components and associated loading maps shown in (a) and (b), respectively. Similar analysis was reported earlier in [59] using our AICrystallographer package (the AtomAI's predecessor).

The prediction of the trained Segmentor for a different experimental image (La-doped BiFeO$_3$) obtained in a different experiment[59] is shown in Figure 4 (b,c) where it was able to remove the experimental noise and separate two sublattices into different classes (red and green in Fig. 4b, c). We note that in this case the input image had a size of 1024 × 1024 pixels while our network was trained only using images of 256 × 256 pixels. This underscores a very important aspect of the fully convolutional neural networks, namely, that they are not sensitive to the size of input image as long as it can be divided by $2^n$, where $n$ is a number of max-pooling layers in the network (here it is equal to 3). However, there is always some optimal pixel-to-angstrom ratio (or, roughly, number of pixels per atom/defect/particle) for which a network will generate the best results. We note that although in this example the Segmentor was trained essentially on a single image, it is generally better to use a large(r) and diverse set of images.

Once we have all atomic coordinates and semantically segmented images, we can perform various forms of multivariate analysis on local image descriptors formed by extracting patches of a fixed size centered around one type of the lattice sites. The results of applying NMF to the Segmentor output are shown in Figure 5. The four NMF components provide clear pictures of atomic displacements whereas the corresponding loading maps show characteristics of the domain structure, all found in an unsupervised manner.

**IV.B. Graph analysis: Localization of topological defects**

In addition to the multivariate analysis of the semantically segmented output, it is also possible to analyze the predicted coordinates using graphs. For example, one can combine a deep learning network for atomic-level semantic segmentation and a depth-first search for traversing a graph formed from predicted coordinates and create a "defect sniffer" that identifies specific types of topological defects (in this example, 5 and 7 member rings) from STEM data on graphene (Figure 6). We note that the semantic segmentation model used in the analysis was trained using the atomic coordinates produced by molecular dynamics simulations of a large variety of structural/topological defects in graphene.

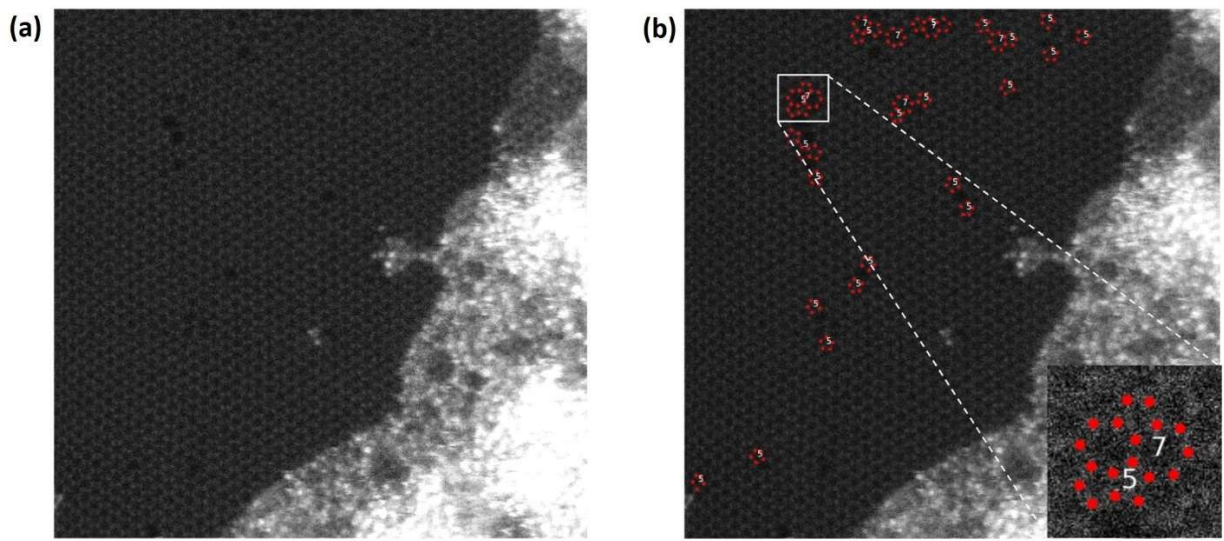

**Figure 6.** (a) Raw experimental STEM image of graphene. (b) Application of the "defect sniffer" (Segmentor + graph analysis) for rapidly (~1 sec) locating specific topological structures (in this case, 5 and 7 member rings) in the image.

## IV.C. Variational Autoencoders (VAE): analysis of structural order parameters

An autoencoder generally refers to a special class of neural networks where the original data set is compressed to a small number of continuous latent variables and is then expanded back to the original data set. In the process, the network learns how to optimally describe the data in terms of the latent variables. This allows the autoencoder to discover the optimal representation(s), while also rejecting the noise present in the data. The variational autoencoder (VAE) builds upon this concept by making the reconstruction process probabilistic. In this case, the latent variables are drawn from a certain (typically standard Gaussian) distribution, and the training process seeks to optimize both the reconstruction loss and the Kullback-Leibler divergence between the encoded distribution and the chosen prior. The advantages of the VAE over classical autoencoders include enforcement of structure and smoothness in the latent space, the ability to learn both continuous and discrete distributions, and state-of-the-art performance for the weakly- and semi-supervised learning (when only a small part of data is labeled). In fact, VAEs have been actively used for discovering trends in various high-dimensional datasets.[127-129] Examples of such trends include emotional expressions in facial databases and writing styles in hand-written digits databases.

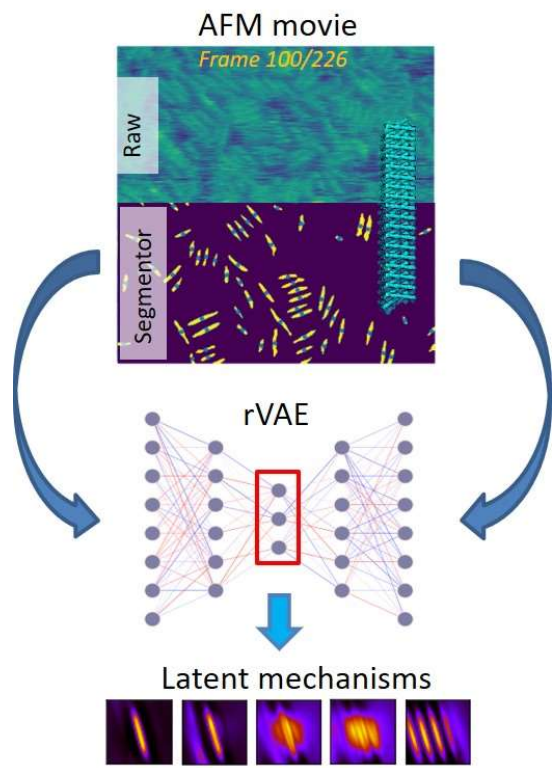

**Figure 7.** Schematics representing the analysis of AFM movie on protein self-organization via the AtomAI's Segmentor and rotationally-invariant variational autoencoder (*r*VAE) for disentangling latent mechanisms associated with local transitions. Based on the results published in Ref [112]

AtomAI expands the classical VAE architecture to disentangle both discrete and continuous representations of the data while accounting for translational and rotational invariances. Recently, the VAE module in AtomAI has been used identify an order parameter in disordered system from atom-resolved movies[60] and to (re)discover molecular building blocks and chemical reaction pathways directly from electron microscopy data in an unsupervised fashion.[130] It has also been used to probe atomic-scale symmetry breaking from CBED patterns.[131]

Here we briefly illustrate the application of the rotationally invariant VAE (*r*VAE) to an atomic force microscopy movie of protein self-organization[112] as shown schematically in Figure 7. We start by applying a pre-trained Segmentor to the AFM movie frames. The Segmentor was trained on just a few labeled images from a "stable" phase of the process characterized by relatively low noise and absence of large scars along the fast scan direction. The Segmentor output then serves as the input (and output) for the *r*VAE which automatically separates the orientation of the particles from other degrees of freedom thereby allowing to encode (and analyze) a rich spectrum of local transitions into the latent space. Depicted in Fig. 7 is a slice of the learned latent manifold associated with an ordering transition. Note that in classical VAE the rotational factor of variation gets admixed into the factors associated with physical mechanisms preventing from learning a proper disentangled representation for any physically meaningful number of latent dimensions.

**IV.D. Predictability of localized functional responses**

The AtomAI's ImSpec models can be trained to predict localized functional responses in the form of 1D spectra from structural 2D images. The method is based on the idea that local structures and functional phenomena are (cor)related and the relationship is parsimonious, that is, it can be explained by a relatively small number of (latent) mechanisms. The ImSpec consists of encoder in the form of convolutional neural network that that maps the 2D images into the low-dimensional latent vector and the decoder in the form of 1D convolutional neural network that reconstructs the spectra from this latent representation. Note that this is different from the (variational) autoencoder approach where inputs and outputs are the same. The correlative structure-property relationships can be analyzed by projecting the learned latent vector representation (for a particular dataset) to the spectral and image domains. As with all supervised ML methods, the caveat is that ImSpec shows good/reliable predictive performance only on the data obtained under similar experimental conditions (see however the next section on some ways of addressing this limitation). In Figure 8, we summarized an example of applying ImSpec-type of models to arrays of plasmonic nanoparticles. Here the inputs are HAADF-STEM structural images whereas the output/targets are EEL spectra. Interestingly, the trained model outputs smoothed spectra closely resembling the shapes of the original ones even though we did not explicitly train it to clean the data. In addition, the analysis of the latent space distributions projected to the image and spectral domains provided an insight into the generative mechanisms of plasmonic interactions in the nanoparticle arrays.[120]

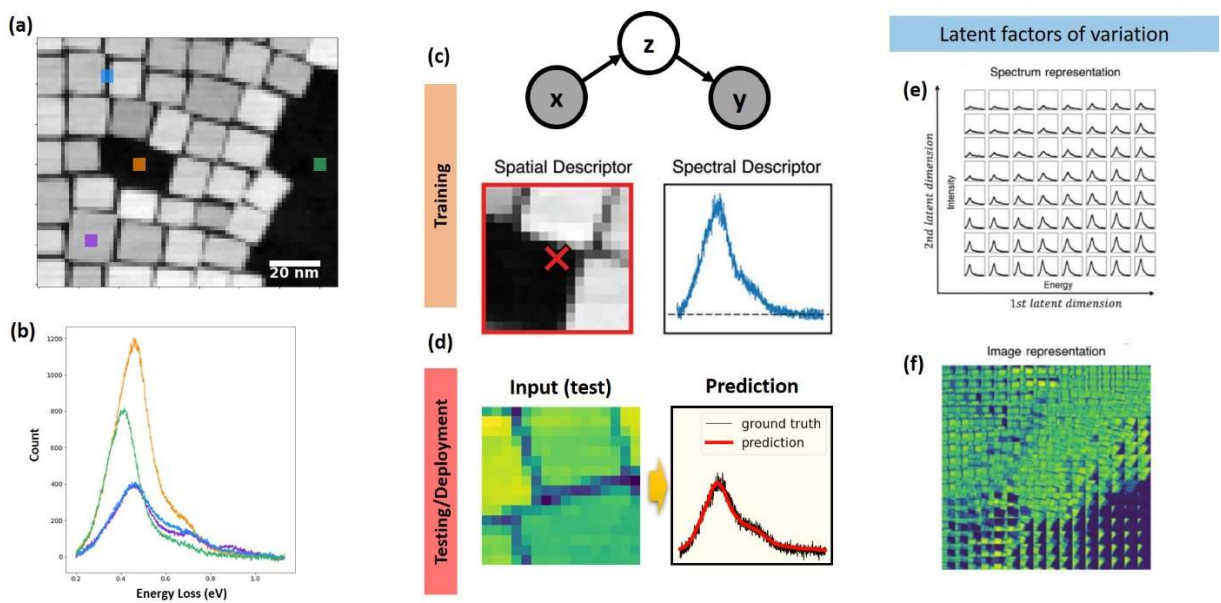

**Figure 8.** Investigating a predictability of plasmonic spectra with an ImSpec model. (a, b) Experimental data: (a) HAADF-STEM image of nanoparticle array and (b) EELS from 4 selected locations in (a). (c) In the ImSpec model, the inputs (**x**) are images and the outputs/targets (**y**) are spectra associated with these images. The correlative structure-property relationship is assumed to be encoded into the latent vector **z** during the model training. (d, e, f) The trained model can be used to predict spectra from new structural images obtained under similar conditions (d) as well

as to view/analyze the latent space representations (e, f) by projecting them image and spectral domains. The images are adapted with changes from Ref [120] (reproduced with permissions from John Wiley and Sons)

### IV.E. Ensemble Learning-Iterative Training (ELIT)

In general, the performance of deep neural networks varies significantly based on the choice of training data and experimental noise, bias, and data acquisition parameters. One way to (partially) account for these issues is to utilize Ensemble Learning (EL) which combines the predictions of multiple models. Here, multiple learners are trained to solve a given problem as compared to commonly constructed ML models that depend on only a single learner. The utilization of a combination of learners allows for a flexibility in the search processes of an algorithm (in this case, a deep learning model), and selection of a broad hypothesis space. Hence, the generalization capability of EL is superior compared to that of a single model, leading to higher prediction accuracies on new data. This approach has already been widely used in various applications[132, 133] ranging from character recognition, text categorization, face recognition to computer-aided medical diagnosis and gene expression analysis.

One such application of interest is feature finding in image data. For image datasets where the tens of thousands of samples with large variabilities are commonly available (such as MNIST and CIFAR), the common deep learning classification strategies work very well. In such cases, the training and validation data are drawn from the (more or less) *i.i.d.* data making the feature-based classifications perform with reasonable accuracies.

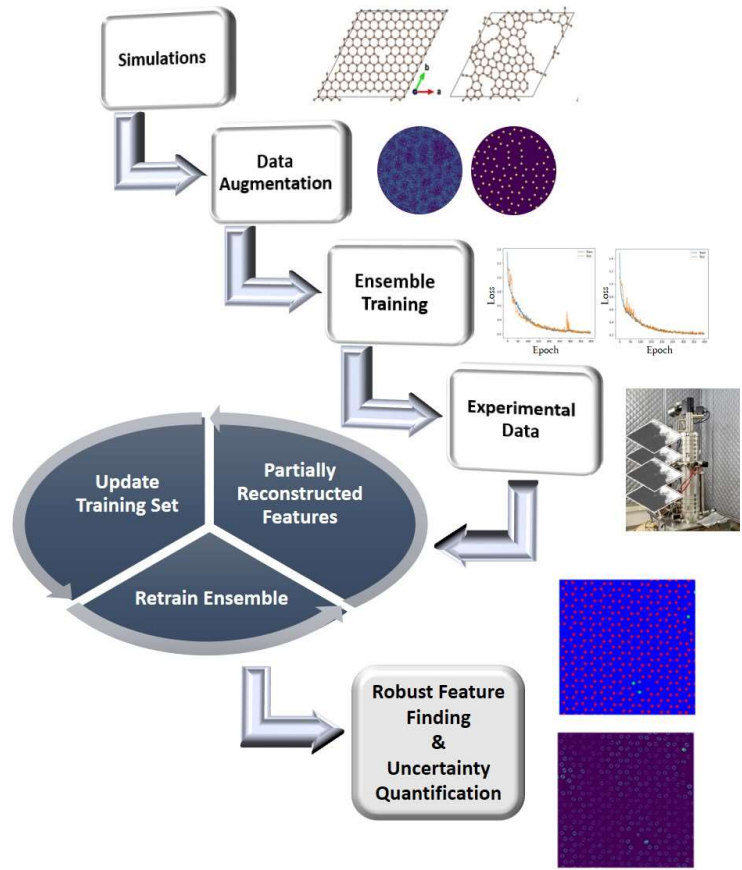

**Figure 9**. An illustrative graphical representation of ELIT workflow[62] demonstrating the main steps of the framework (the photo of a microscope is obtained from the ORNL photo gallery with permission for public release)

In contrast, for datasets consisting of atomically resolved STEM or STM images, the presence of almost identical atoms or objects is common, even when the imaging or simulation conditions are varied. The disparities in detected features for many experimental or atomistic simulation conditions are often negligible which makes the task of feature finding much more challenging. As a result, deep neural networks trained to account for a large variety of experimental conditions are not always good enough to recognize subtle distinctions in atomic features present in a particular experimental dataset. In addition, the applications of deep neural networks to such problems must be able to deal with out-of-distribution effects by rapidly adapting to changes in the imaging conditions.

Ensemble learning combined with iterative training (ELIT) offers a strategy to surpass the above-mentioned challenges (Figure 9). The EL part of the workflow allows for selection of artifact-free features and pixel-wise uncertainty maps by combining multiple networks. The IT part retrains the ensemble networks with already detected features, focusing its attention on features present in the (heavily degenerate) data and thus increasing the detection limit of the network on the dataset(s) of interest.

We have recently successfully applied a ELIT framework to multiple datasets, including the dynamic STEM data from graphene with impurities and static STEM data from NiO-LSMO.[62] In both cases, the learning begins with training the models on data generated by simulations and consequently applying it to the experimental images followed by iterative reevaluations and modifications to the training sets to obtain a model that can successfully identify all atoms and impurities, or defects present in the system. The ELIT framework works reasonably well for feature finding in new experimental data and allows a generation of meaningful uncertainty maps on the level of individual pixels. This method also allows for rapid correction for out of distribution effects during automated experiments where the imaging conditions change compared to the training set.

**V. CONNECTION TO AB-INITIO SIMULATIONS**

As described in the previous sections, AtomAI allows using deep fully convolutional neural networks (Segmentor models) for finding atoms and/or other relevant features (particles, defects, etc.) from microscopic images. The output contains a list of coordinates with corresponding classes for all detected atoms, which can easily be converted to Cartesian or fractional coordinates as necessary. We note that one of the key steps before performing any atomistic simulation is to build the system's "cell". It can be of different types such as bulk conventional unit cell, supercell or surface depending on the type of simulations and material properties of interest. Hence, once we obtain the coordinates corresponding to all atoms present in the system from the Segmentor model, our next step is to construct one of these cells.

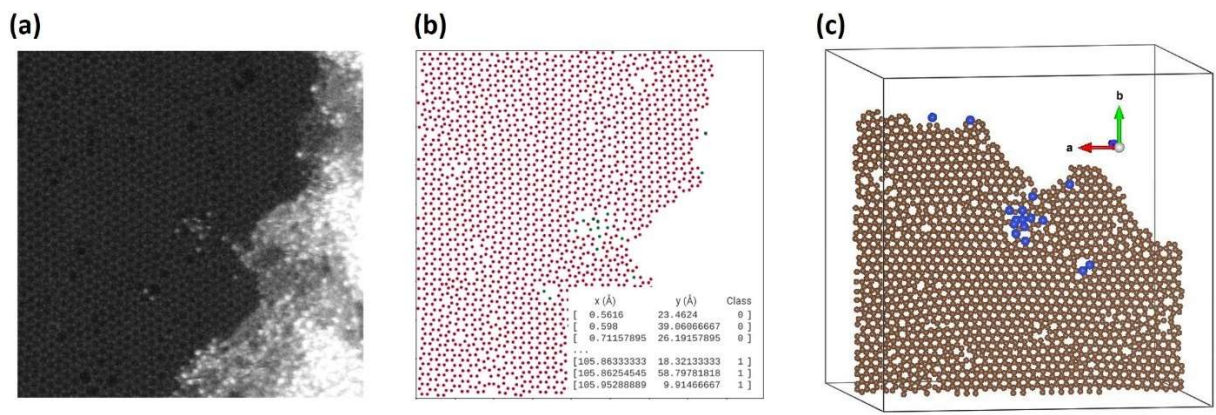

**Figure 10.** Schematic of (a) experimental STEM image of graphene, (b) atomic types and coordinates predicted by AtomAI's Segmentor and (c) a corresponding supercell to be used for performing atomistic simulations.

Figure 10 shows an example of a STEM image of graphene, the AtomAI's Segmentor-predicted coordinates for C and Si atoms, and the corresponding supercell built using the AtomAI's `ase_obj_basic` function. Once these cells are constructed, the next task is to optimize the geometry. One can obtain the relaxed geometry using quasi-Newton algorithms or first-principles methods before proceeding to perform other advanced simulations such as computing the electronic properties.

Within the quasi-Newton methods as implemented in ASE, the algorithms decide the positions of atoms at every iteration based on the forces and second derivative of the total energy of the atoms. This may not work well in all cases, where electronic interactions can play a major role to acquire relaxed geometry. In such cases, we can employ the objects obtained using AtomAI utility functions in electronic structure codes to first relax atom positions, cell shape, and volume such that forces and stresses between all atoms are negligible. Consequently, atomistic simulations with various classical potentials such Lennard-Jones using built-in ASE calculators or other ab-initio DFT (using pseudopotentials) or MD codes can be employed to evaluate material properties.

## VI. CONCLUSIONS

The AtomAI package is introduced as a high-level package bridging the instrument-specific Python-based ecosystem and deep learning and simulation tools into a single workflow. AtomAI allows direct applications of the deep convolutional neural networks for atomic and mesoscopic image segmentation, converting the image and spectroscopy data into class-based descriptors. For atomically-resolved data, these include the type and positions of atomic species. AtomAI further allows the implementation of a broad range of image and spectrum analysis classes and methods, including simple and rotationally/translationally invariant variational autoencoders for unsupervised classification and disentanglement of continuous traits in datasets. These concepts are further extended to im2spec/spec2im type of models for predicting functional spectra from structural images and vice versa. Finally, AtomAI allows seamless connection to the Python-based first-principles modeling including molecular dynamics and density-functional theory models using the derived atomic position.

While the majority of applications to date were based on the atomically resolved STEM, the flexibility of AtomAI allows straightforward extension towards the analysis of mesoscopic imaging data once the labels and feature identification workflows are available.

**Acknowledgements:** This effort was performed and partially supported (M.Z.) at the Oak Ridge National Laboratory's Center for Nanophase Materials Sciences (CNMS), a U.S. Department of Energy, Office of Science User Facility, and by U.S. Department of Energy, Office of Science, Office of Basic Energy Sciences Data, Artificial Intelligence and Machine Learning at DOE Scientific User Facilities program under the Digital Twin Project (Award Number 34532) (A.G.) and MLExchange Project (Award Number 107514) (T.W., S.V.K.) The authors gratefully acknowledge multiple discussions with Matt Chisholm, Andy Lupini, Mark Oxley, Kevin Roccapriore, Jordan Hachtel, Ondrej Dyck and multiple other colleagues at ORNL whose advice and beta testing have been instrumental throughout the development of AtomAI from 2019 to 2021 and its predecessor AICrystallographer in 2016 -2019. The authors also express their deep gratitude to Colin Ophus (LBNL), Steven Spurgeon (PNNL), Francisco de la Peña (University of Lille), Dieter Weber (Juelich), and Ian Maclaren (Glasgow University) for critical reading of the manuscript, suggesting several key references, and suggesting improvement of key figures.

**Uncategorized References**

1. S. J. Pennycook, Ultramicroscopy **123**, 28-37 (2012).
2. S. J. Pennycook and P. D. Nellist, (Springer, New York, 2011).
3. N. Dellby, O. L. Krivanek, P. D. Nellist, P. E. Batson and A. R. Lupini, J. Electron Microsc. **50** (3), 177-185 (2001).
4. P. Y. Huang, S. Kurasch, J. S. Alden, A. Shekhawat, A. A. Alemi, P. L. McEuen, J. P. Sethna, U. Kaiser and D. A. Muller, Science **342** (6155), 224-227 (2013).
5. C. Ophus, Microsc. microanal. **25** (3), 563-582 (2019).
6. M. v. Ardenne, Z. Tech. Phys **19**, 407-416 (1938).
7. M. v. Ardenne, Zeit. Physik **109**, 553-572 (1938).
8. M. Knoll and E. Ruska, Zeit. Physik **78** (5), 318-339 (1932).
9. E. Ruska, Rev. Mod. Phys. **59** (3), 627-638 (1987).
10. G. Binnig, H. Rohrer, C. Gerber and E. Weibel, Physical Review Letters **50** (2), 120-123 (1983).
11. G. Binnig, C. F. Quate and C. Gerber, Phys. Rev. Lett. **56** (9), 930-933 (1986).
12. R. Garcia and R. Perez, Surf. Sci. Rep. **47** (6-8), 197-301 (2002).
13. A. Gruverman, O. Auciello, R. Ramesh and H. Tokumoto, Nanotechnology **8**, A38-A43 (1997).
14. Y. Martin and H. K. Wickramasinghe, Appl. Phys. Lett. **50** (20), 1455-1457 (1987).
15. O. Vatel and M. Tanimoto, J. Appl. Phys. **77** (6), 2358-2362 (1995).
16. M. Nonnenmacher, M. P. Oboyle and H. K. Wickramasinghe, Appl. Phys. Lett. **58** (25), 2921-2923 (1991).
17. O. L. Krivanek, N. Dellby, A. J. Spence, R. A. Camps and L. M. Brown, in *Electron Microscopy and Analysis 1997*, edited by J. M. Rodenburg (Iop Publishing Ltd, Bristol, 1997), pp. 35-40.
18. P. E. Batson, N. Dellby and O. L. Krivanek, Nature **418** (6898), 617-620 (2002).
19. M. Varela, S. D. Findlay, A. R. Lupini, H. M. Christen, A. Y. Borisevich, N. Dellby, O. L. Krivanek, P. D. Nellist, M. P. Oxley, L. J. Allen and S. J. Pennycook, Physical Review Letters **92** (9), 095502 (2004).
20. W. Zhou, X. L. Zou, S. Najmaei, Z. Liu, Y. M. Shi, J. Kong, J. Lou, P. M. Ajayan, B. I. Yakobson and J. C. Idrobo, Nano Lett. **13** (6), 2615-2622 (2013).
21. N. D. Browning, J. P. Buban, P. D. Nellist, D. P. Norton, M. F. Chisholm and S. J. Pennycook, Physica C **294** (3-4), 183-193 (1998).
22. X. Guo and R. Waser, Progress in Materials Science **51** (2), 151-210 (2006).
23. B. J. McMorran, A. Agrawal, I. M. Anderson, A. A. Herzing, H. J. Lezec, J. J. McClelland and J. Unguris, Science **331** (6014), 192-195 (2011).
24. V. Grillo, E. Karimi, G. C. Gazzadi, S. Frabboni, M. R. Dennis and R. W. Boyd, Phys. Rev. X **4** (1), 7 (2014).
25. J. Rusz and S. Bhowmick, Phys. Rev. Lett. **111** (10), 5 (2013).
26. J. C. Idrobo, A. R. Lupini, T. L. Feng, R. R. Unocic, F. S. Walden, D. S. Gardiner, T. C. Lovejoy, N. Dellby, S. T. Pantelides and O. L. Krivanek, Phys. Rev. Lett. **120** (9) (2018).
27. S. H. Cho, K. M. Roccapriore, C. K. Dass, S. Ghosh, J. Choi, J. Noh, L. C. Reimnitz, S. Heo, K. Kim, K. Xie, B. A. Korgel, X. Q. Li, J. R. Hendrickson, J. A. Hachtel and D. J. Milliron, J. Chem. Phys. **152** (1), 17 (2020).
28. R. Senga, K. Suenaga, P. Barone, S. Morishita, F. Mauri and T. Pichler, Nature **573** (7773), 247-+ (2019).
29. R. F. Egerton, Ultramicroscopy **107** (8), 575-586 (2007).
30. Y. Jiang, Z. Chen, Y. M. Hang, P. Deb, H. Gao, S. E. Xie, P. Purohit, M. W. Tate, J. Park, S. M. Gruner, V. Elser and D. A. Muller, Nature **559** (7714), 343-+ (2018).
31. N. Shibata, T. Seki, G. Sanchez-Santolino, S. D. Findlay, Y. Kohno, T. Matsumoto, R. Ishikawa and Y. Ikuhara, Nat. Commun. **8**, 7 (2017).

32. Y. S. Yang, C. C. Chen, M. C. Scott, C. Ophus, R. Xu, A. Pryor, L. Wu, F. Sun, W. Theis, J. H. Zhou, M. Eisenbach, P. R. C. Kent, R. F. Sabirianov, H. Zeng, P. Ercius and J. W. Miao, Nature **542** (7639), 75-+ (2017).
33. C. L. Jia, V. Nagarajan, J. Q. He, L. Houben, T. Zhao, R. Ramesh, K. Urban and R. Waser, Nature Materials **6** (1), 64-69 (2007).
34. C. L. Jia, S. B. Mi, K. Urban, I. Vrejoiu, M. Alexe and D. Hesse, Nature Materials **7** (1), 57-61 (2008).
35. M. F. Chisholm, W. D. Luo, M. P. Oxley, S. T. Pantelides and H. N. Lee, Physical Review Letters **105** (19) (2010).
36. I. MacLaren and Q. M. Ramasse, International Materials Reviews **59** (3), 115-131 (2014).
37. C. L. Jia, S. B. Mi, M. Faley, U. Poppe, J. Schubert and K. Urban, Physical Review B **79** (8) (2009).
38. A. Borisevich, O. S. Ovchinnikov, H. J. Chang, M. P. Oxley, P. Yu, J. Seidel, E. A. Eliseev, A. N. Morozovska, R. Ramesh, S. J. Pennycook and S. V. Kalinin, ACS Nano **4** (10), 6071-6079 (2010).
39. Y. M. Kim, A. Kumar, A. Hatt, A. N. Morozovska, A. Tselev, M. D. Biegalski, I. Ivanov, E. A. Eliseev, S. J. Pennycook, J. M. Rondinelli, S. V. Kalinin and A. Y. Borisevich, Adv. Mater. **25** (17), 2497-2504 (2013).
40. Q. He, R. Ishikawa, A. R. Lupini, L. Qiao, E. J. Moon, O. Ovchinnikov, S. J. May, M. D. Biegalski and A. Y. Borisevich, ACS Nano **9** (8), 8412-8419 (2015).
41. M. Nord, A. Ross, D. McGrouther, J. Barthel, M. Moreau, I. Hallsteinsen, T. Tybell and I. MacLaren, Physical Review Materials **3** (6), 063605 (2019).
42. Y. M. Kim, A. Morozovska, E. Eliseev, M. P. Oxley, R. Mishra, S. M. Selbach, T. Grande, S. T. Pantelides, S. V. Kalinin and A. Y. Borisevich, Nat. Mater. **13** (11), 1019-1025 (2014).
43. A. Y. Borisevich, A. R. Lupini, J. He, E. A. Eliseev, A. N. Morozovska, G. S. Svechnikov, P. Yu, Y. H. Chu, R. Ramesh, S. T. Pantelides, S. V. Kalinin and S. J. Pennycook, Phys. Rev. B **86** (14) (2012).
44. A. B. Yankovich, B. Berkels, W. Dahmen, P. Binev, S. I. Sanchez, S. A. Bradley, A. Li, I. Szlufarska and P. M. Voyles, Nature Communications **5** (2014).
45. L. Vlcek, M. Ziatdinov, A. Maksov, A. Tselev, A. P. Baddorf, S. V. Kalinin and R. K. Vasudevan, ACS Nano **13** (1), 718-727 (2019).
46. L. Vlcek, A. Maksov, M. H. Pan, R. K. Vasudevan and S. V. Kahnin, Acs Nano **11** (10), 10313-10320 (2017).
47. W. Z. Lin, Q. Li, A. Belianinov, B. C. Sales, A. Sefat, Z. Gai, A. P. Baddorf, M. H. Pan, S. Jesse and S. V. Kalinin, Nanotechnology **24** (41) (2013).
48. Z. Gai, W. Z. Lin, J. D. Burton, K. Fuchigami, P. C. Snijders, T. Z. Ward, E. Y. Tsymbal, J. Shen, S. Jesse, S. V. Kalinin and A. P. Baddorf, Nat. Commun. **5** (2014).
49. O. S. Ovchinnikov, S. Jesse, P. Bintacchit, S. Trolier-McKinstry and S. V. Kalinin, Physical Review Letters **103** (15) (2009).
50. A. Kumar, O. Ovchinnikov, S. Guo, F. Griggio, S. Jesse, S. Trolier-McKinstry and S. V. Kalinin, Phys. Rev. B **84** (2), 024203 (2011).
51. M. P. Nikiforov, V. V. Reukov, G. L. Thompson, A. A. Vertegel, S. Guo, S. V. Kalinin and S. Jesse, Nanotechnology **20** (40), 405708 (2009).
52. S. Jesse and S. V. Kalinin, Nanotechnology **20** (8), 085714 (2009).
53. M. Bosman, M. Watanabe, D. T. L. Alexander and V. J. Keast, Ultramicroscopy **106** (11-12), 1024-1032 (2006).
54. L. J. Allen, A. J. D'Alfonso, S. D. Findlay, M. P. Oxley, M. Bosman, V. J. Keast, E. C. Cosgriff, G. Behan, P. D. Nellist and A. I. Kirkland, in *Electron Microscopy and Multiscale Modeling, Proceedings*, edited by A. S. Avilov, S. L. Dudarev and L. D. Marks (Amer Inst Physics, Melville, 2008), Vol. 999, pp. 32-46.
55. R. Kannan, A. V. Ievlev, N. Laanait, M. A. Ziatdinov, R. K. Vasudevan, S. Jesse and S. V. Kalinin, Adv. Struct. Chem. Imag. **4**, 20 (2018).

56. R. K. Vasudevan, M. Ziatdinov, S. Jesse and S. V. Kalinin, Nano Letters **16** (9), 5574-5581 (2016).
57. A. Paszke, S. Gross, F. Massa, A. Lerer, J. Bradbury, G. Chanan, T. Killeen, Z. Lin, N. Gimelshein and L. Antiga, Advances in neural information processing systems, 8026-8037 (2019).
58. L. Buitinck, G. Louppe, M. Blondel, F. Pedregosa, A. Mueller, O. Grisel, V. Niculae, P. Prettenhofer, A. Gramfort and J. Grobler, arXiv preprint arXiv:1309.0238 (2013).
59. M. Ziatdinov, C. Nelson, R. K. Vasudevan, D. Y. Chen and S. V. Kalinin, Appl. Phys. Lett. **115** (5), 5 (2019).
60. S. V. Kalinin, O. Dyck, S. Jesse and M. Ziatdinov, Science Advances **7** (17), eabd5084 (2021).
61. M. Ziatdinov, S. Jesse, B. G. Sumpter, S. V. Kalinin and O. Dyck, Nanotechnology **32** (3), 035703 (2020).
62. A. Ghosh, B. G. Sumpter, O. Dyck, S. V. Kalinin and M. Ziatdinov, arXiv preprint arXiv:2101.08449 (2021).
63. M. Ziatdinov, U. Fuchs, J. H. Owen, J. N. Randall and S. V. Kalinin, arXiv preprint arXiv:2002.04716 (2020).
64. S. V. Kalinin, J. J. Steffes, B. D. Huey and M. Ziatdinov, arXiv preprint arXiv:2007.06194 (2020).
65. R. Ignatans, M. Ziatdinov, R. Vasudevan, M. Valleti, V. Tileli and S. V. Kalinin, arXiv preprint arXiv:2011.11869 (2020).
66. S. V. Kalinin, A. R. Lupini, O. Dyck, S. Jesse, M. Ziatdinov and R. K. Vasudevan, MRS Bull. **44** (7), 565-575 (2019).
67. F. de la Peña, T. Ostasevicius, V. T. Fauske, P. Burdet, P. Jokubauskas, M. Nord, M. Sarahan, E. Prestat, D. N. Johnstone, J. Taillon and Microanalysis, Microscopy and Microanalysis **23** (S1), 214-215 (2017).
68. S. Somnath, C. R. Smith, N. Laanait, R. K. Vasudevan and S. Jesse, Microsc. microanal. **25** (S2), 220-221 (2019).
69. D. Mukherjee, GitHub repository, https://github.com/pycroscopy/stemtool (2020).
70. G. Duscher, GitHub repository, https://github.com/pycroscopy/pyTEMlib (2020).
71. D. Johnstone, GitHub Repository, https://pyxem.github.io/pyxem-website/.
72. A. Clausen, D. Weber, K. Ruzaeva, V. Migunov, A. Baburajan, A. Bahuleyan, J. Caron, R. Chandra, S. Halder and M. Nord, Journal of Open Source Software **5** (50), 2006 (2020).
73. B. H. Savitzky, L. Hughes, K. C. Bustillo, H. D. Deng, N. L. Jin, E. G. Lomeli, W. C. Chueh, P. Herring, A. Minor and C. Ophus, Microsc. microanal. **25** (S2), 124-125 (2019).
74. I. MacLaren, T. A. Macgregor, C. S. Allen and A. I. Kirkland, APL Materials **8** (11), 110901 (2020).
75. L. J. Allen, A. J. D'Alfonso and S. D. Findlay, Ultramicroscopy **151**, 11-22 (2015).
76. C. T. Koch, Ph. D. Thesis (2002).
77. J. Madsen and T. Susi, Microsc. microanal. **26** (S2), 448-450 (2020).
78. I. Lobato and D. Van Dyck, Ultramicroscopy **156**, 9-17 (2015).
79. J. O. Oelerich, L. Duschek, J. Belz, A. Beyer, S. D. Baranovskii and K. Volz, Ultramicroscopy **177**, 91-96 (2017).
80. A. Pryor, C. Ophus and J. Miao, Adv. Struct. Chem. Imag. **3** (1), 15 (2017).
81. J. Barthel, Ultramicroscopy **193**, 1-11 (2018).
82. A. G. e. al., Journal of Chemical Physics **152**, 204108.
83. P. G. e. al., Journal of Physics: Condensed Matter **21**, 395502 (2009).
84. P. G. e. al., Journal of Physics: Condensed Matter **29(46)**, 465901 (2017).
85. P. G. e. al., J. Chem. Phys **152**, 154105 (2020).
86. T. D. K. e. al., Journal of Chemical Physics **152**, 194103 (2020).
87. D. R. G. C.J. Permann, D. Andrš, R. W. Carlsen, F. Kong, A. D. Lindsay, J. M. Miller, J. W. Peterson, A. E. Slaughter, R. H. Stogner and R. C. Martineau, SoftwareX **11**.

88. D. Guido, *The Finite Element Method for Three-Dimensional Thermomechanical Applications*. (John Wiley & Sons, 2004).
89. https://lammps.sandia.gov/.
90. https://www.3ds.com/products-services/biovia/products/molecular-modeling-simulation/biovia-materials-studio/.
91. E. A. J.M. Soler, J. D. Gale, A. García, J. Junquera, P. Ordejón and D. Sánchez-Portal, Journal of Physics: Condensed Matter **14**, 2745 (2002).
92. A. S. G. K. C. Koppenhoefer, C. Ruggieri, R. H. Dodds, Robert H. Jr and B. E. Healey, Report No. Civil Engineering Studies SRS-596, 1994.
93. P. M. Malinen and P. Råback, Appl. Mater. Sci. **19**, 101-113 (2013).
94. C. G. C. Kloss, A. Hager, S. Amberger, S. Pirker, Progress in Computational Fluid Dynamics **12**, 140-152 (2012).
95. W. L. DeLano, CCP4 Newsletter on protein crystallography **40**, 82-92 (2002).
96. P. Hirel, Comput. Phys, Comm. **197**, 212-219 (2015).
97. K. M. a. F. Izumi, Journal of applied crystallography **44(6)**, 1272-1276 (2011).
98. B. G. J. Ahrens, C. Law, (Elsevier, 2015).
99. R. A. L. Martínez, E. G. Birgin, J. M. Martínez. , Journal of Computational Chemistry **30(13)**, 2157-2164 (2009).
100. D. E. C. M.D. Hanwell, D.C. Lonie, T. Vandermeersch, E. Zurek and G. R. Hutchison Journal of Cheminformatics **4(1)**, 1-17 (2012).
101. A. Stukowski, Modelling Simul. Mater. Sci. Eng. **18**, 015012 (2010).
102. A. Utkarsh, (Kitware, 2015).
103. A. D. W. Humphrey, and K. Schulten, J. Molec. Graphics **14**, 33-38 (1996).
104. S. P. O. A. Jain, G. Hautier, W. Chen, W.D. Richards, S. Dacek, S. Cholia, D. Gunter, D. Skinner, G. Ceder, K.A. Persson, APL Materials **1** (1) (2013).
105. M. J. M. D. Hicks, E. Gossett, C. Toher, O. Levy, R. M. Hanson, G. L. W. Hart, and S. Curtarolo, Comp. Mat. Sci. **161**, S1-S1011 (2019).
106. S. K. J. E. Saal, M. Aykol, B. Meredig, and C. Wolverton, JOM **65**, 1501-1509 (2013).
107. D. H. M. J. Mehl, C. Toher, O. Levy, R. M. Hanson, G. L. W. Hart, and S. Curtarolo, Comp. Mat. Sci. **136**, S1-S828 (2017).
108. S. J. E. S. S. Kirklin, B. Meredig, B, A. Thompson, J.W. Doak, M. Aykol, S. Rühl and C. Wolverton. , npj Computational Materials **1** (2015).
109. C. D. a. M. Scheffler, J. Phys. Mater **2**, 036001 (2019).
110. S. Somnath, G. Duscher, M. Ziatdinov, R. Vasudevan and M. Valleti, GitHub repository, 10.5281/zenodo.4679761 (2021).
111. M. Ziatdinov, S. Zhang, O. Dollar, J. Pfaendtner, C. J. Mundy, X. Li, H. Pyles, D. Baker, J. J. De Yoreo and S. V. Kalinin, Nano Letters **21** (1), 158-165 (2021).
112. S. V. Kalinin, S. Zhang, M. Valleti, H. Pyles, D. Baker, J. J. De Yoreo and M. Ziatdinov, ACS Nano **15** (4), 6471-6480 (2021).
113. M. Ziatdinov, O. Dyck, X. Li, B. G. Sumpter, S. Jesse, R. K. Vasudevan and S. V. Kalinin, Science advances **5** (9), eaaw8989 (2019).
114. O. Ronneberger, P. Fischer and T. Brox, in *International Conference on Medical image computing and computer-assisted intervention* (Springer, Cham, 2015), pp. 234-241.
115. S. Xie and Z. Tu, Proceedings of the IEEE international conference on computer vision, 1395-1403 (2015).
116. K. He, X. Zhang, S. Ren and J. Sun, Proceedings of the IEEE conference on computer vision and pattern recognition, 770-778 (2016).

117. V. Badrinarayanan, A. Kendall and R. Cipolla, IEEE transactions on pattern analysis **39** (12), 2481-2495 (2017).
118. A. G. Wilson and P. Izmailov, arXiv preprint arXiv:2002.08791 (2020).
119. S. V. Kalinin, K. Kelley, R. K. Vasudevan and M. Ziatdinov, ACS Appl. Mater. Interfaces **13** (1), 1693-1703 (2021).
120. K. M. Roccapriore, M. Ziatdinov, S. H. Cho, J. A. Hachtel and S. V. Kalinin, Small **n/a** (n/a), 2100181 (2021).
121. B. Lakshminarayanan, A. Pritzel and C. Blundell, in *Proceedings of the 31st International Conference on Neural Information Processing Systems* (Curran Associates Inc., Long Beach, California, USA, 2017), pp. 6405–6416.
122. M. Ziatdinov, N. Creange, X. Zhang, A. Morozovska, E. Eliseev, R. K. Vasudevan, I. Takeuchi, C. Nelson and S. V. Kalinin, Appl. Phys. Rev. **8** (1), 011403 (2021).
123. R. K. Vasudevan, K. Kelley, H. Funakubo, S. Jesse, S. V. Kalinin and M. Ziatdinov, arXiv preprint arXiv:2011.13050 (2020).
124. D. P. Kingma and M. Welling, Foundations and Trends® in Machine Learning **12** (4), 307-392 (2019).
125. M. Ziatdinov, GitHub repository, https://github.com/pycroscopy/atomai (2020).
126. C. Nelson, A. Ghosh, M. Ziatdinov and S. Kalinin V, (Zenodo, 2021).
127. D. Celis and M. Rao, in *Proceedings of the 1st International Workshop on Fairness, Accountability, and Transparency in MultiMedia* (Association for Computing Machinery, Nice, France, 2019), pp. 26–32.
128. T. Yamada, M. Hosoe, K. Kato and K. Yamamoto, presented at the 2017 14th IAPR International Conference on Document Analysis and Recognition (ICDAR), 2017 (unpublished).
129. C. Doersch, **2021**, arXiv:1606.05908. arXiv.org e-Print archive. https://arxiv.org/abs/1606.05908.
130. S. V. Kalinin, O. Dyck, A. Ghosh, B. G. Sumpter and M. Ziatdinov, arXiv preprint arXiv:2010.09196 (2020).
131. M. P. Oxley, M. Ziatdinov, O. Dyck, A. R. Lupini, R. Vasudevan and S. V. Kalinin, npj Computational Materials **7** (1), 65 (2021).
132. G. S. a. J. Elder, *Ensemble Methods in Data Mining: Improving Accuracy Through Combining Predictions*. (Morgan and Claypool Publishers, 2010).
133. C. Z. a. Y. Ma, *Ensemble machine learning: methods and applications*. (Springer Science & Business Media, 2012).